\documentclass{article}
\usepackage{epsfig,amsmath,amsthm,amssymb,graphics,verbatim,amsfonts,psfrag}

\usepackage{makeidx}

\theoremstyle{plain}
\newtheorem{thm}{Theorem}[section]
\newtheorem{prop}[thm]{Proposition}

\theoremstyle{definition}

\def\R{\mathbb{R}}

\def\Z{\mathbb{Z}}

\def\I{\infty}
\def\ra{\rightarrow}

\newcommand{\be}{\begin{equation}}
\newcommand{\ee}{\end{equation}}
\newcommand{\benn}{\begin{equation*}}
\newcommand{\eenn}{\end{equation*}}
\newcommand{\bea}{\begin{eqnarray}}
\newcommand{\eea}{\end{eqnarray}}
\newcommand{\beann}{\begin{eqnarray*}}
\newcommand{\eeann}{\end{eqnarray*}}

\begin{document}
 
\date{}
\title{Complex Eigenvalues for Binary Subdivision Schemes}
\author{Christian K\"{u}hn}

\maketitle

\begin{center}

\small{Cornell University, Center for Applied Mathematics, 14853 Ithaca NY, United States}\\
\small{e-mail: ck274@cornell.edu}
\end{center}

\begin{abstract}

Convergence properties of binary stationary subdivision schemes for curves have been analyzed using the techniques of z-transforms and eigenanalysis. Eigenanalysis provides a way to determine derivative continuity at specific points based on the eigenvalues of a finite matrix. None of the well-known subdivision schemes for curves have complex eigenvalues. We prove when a convergent scheme with palindromic mask can have complex eigenvalues and that a lower limit for the size of the mask exists in this case. We find a scheme with complex eigenvalues achieving this lower bound. Furthermore we investigate this scheme numerically and explain from a geometric viewpoint why such a scheme has not yet been used in computer-aided geometric design. 

\end{abstract}


\section{Introduction and Notation}

Subdivision schemes have been used successfully in computer-aided geometric design for more than two decades. The basic idea is to consider a mesh of points in either one or two dimensions and to construct a curve or a surface out of this mesh by an iterative procedure. This allows the representation of geometric objects by a relatively small number of points. In this paper we only consider subdivision schemes for curves. In the following paragraph, we review the required background very briefly and refer to \cite{5} and \cite{6} for a more detailed introduction.\\

\begin{figure}[htbp]
\psfrag{primal}{primal}
\psfrag{dual}{dual}
	\centering
		\includegraphics[width=0.80\textwidth]{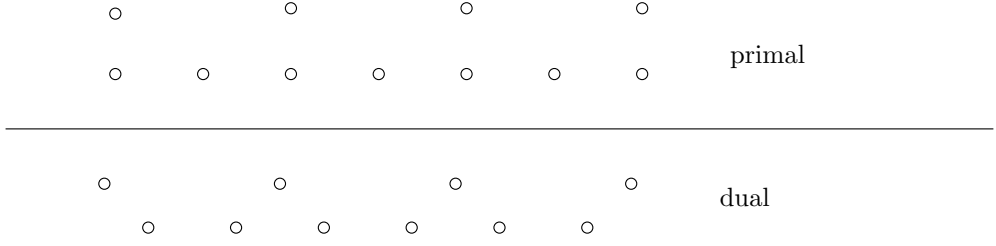}
	\caption{Difference between primal and dual refinement of the mesh} 
	\label{fig:fig2}
\end{figure}

We start with an inital mesh of points $N_k=2^{-k}\mathbb{Z}$ and associate with it a sequence of control points $\textbf{P}^k=\{P^k_i\}$. As an example consider the case $k=1$; then we can view the previous construction as a bi-infinite sequence of points
\benn
 \{\ldots,(-1,P^k_{-2}),(-1/2,P^k_{-1}),(0,P^k_{0}),(1/2,P^k_{1}),\ldots\} 
\eenn
A binary subdivision scheme doubles the number of points at each step, obtaining all new points only using the points of the last step. The mesh $N_k$ is refined to $N_{k+1}$ for primal schemes and to a mesh $\{2^{-(k-1)}\mathbb{Z}-2^{-(k-2)}\mathbb{Z}\}$ for dual schemes (see Figure \ref{fig:fig2}).\\

To simplify notation we simply suppress this underlying refinement of the mesh and focus on the construction of the sequence of control points $\textbf{P}^k$. The main data defining a binary stationary subdivision scheme $S_{\textbf{a}}$ is given by a so-called mask $\textbf{a}=\{a_i\in\R|i\in\sigma(\textbf{a}) \subset \Z \}$ and refinement rules
\be
\label{eq:rules}
P^{k+1}_{2i}=\sum_{2j\in\sigma(A)}a_{2j}P^k_{i-j}\quad \text{ and } \quad P^{k+1}_{2i+1}=\sum_{2j+1\in\sigma(A)}a_{2j+1}P^k_{i-j}
\ee
where $\sigma(\textbf{a})$ is called the support of $S_{\textbf{a}}$. The iteration of a subdivision scheme is denoted by $\textbf{P}^{k+1}=S_{\textbf{a}}\textbf{P}^k$. Furthermore we consider only palindromic masks with finite width, i.e. we suppose $\sigma(\textbf{a})=\{-n,-n+1,\cdots,n-1,n\}$ and $a_{i}=a_{-i}$ if $n$ is odd (primal schemes) and $\sigma(\textbf{a})=\{-n+1,\cdots,n-1,n\}$ and $a_{i}=a_{1-i}$ if $n$ is even (dual schemes).\\

The next natural step is to analyze the limiting properties of a scheme as $k\ra \I$. Let $p^k(t)$ denote the function obtained from the control points $\textbf{P}^k$ (attached to the underlying mesh) by linear interpolation. A subdivision scheme $S_{\textbf{a}}$ is termed $C^m$-convergent if there exists a $C^m$ function $f:\R\rightarrow \R$ such that 
\benn
\lim_{k\rightarrow\I}\max_{A\leq t\leq B}\|f(t)-p^k(t)\|=0
\eenn
for every fixed $A,B\in\R$ and if $f(t)\not\equiv 0$ for some initial sequence of control points. Two paradigms of analysis have been developed to answer whether a scheme $S_{\textbf{a}}$ is convergent. First, we define the z-transform of a subdivision scheme by
\be
\label{eq:ztrans}
S_{\textbf{a}} \quad \longleftrightarrow \quad s_{\textbf{a}}(z)=\sum_{i\in\sigma(\textbf{a})}a_{i}z^{i}
\ee
giving a one-to-one correspondence between the mask $\textbf{a}$ and a formal Laurent series. We have the following well-known results (see \cite{1} or \cite{2}):
\begin{thm}
\label{thm:nec}
$S_{\textbf{a}}$ is not convergent if $s_{\textbf{a}}(1)\neq 2$ or $s_{\textbf{a}}(-1)\neq 0$.
\end{thm}
Hence convergent schemes have $s_{\textbf{a}}(z)=(1+z)s_{\textbf{b}}(z)$ where the scheme $S_{\textbf{b}}$ relates to $S_{\textbf{a}}$ via $\Delta(S_{\textbf{a}}\textbf{P}^k)=S_{\textbf{b}}\Delta\textbf{P}^k$ where $\Delta\textbf{P}^k=\{P^k_{i}-P^{k-1}_{i}|i\in\Z\}$. We call $S_{\textbf{b}}$ the difference scheme associated to $S_{\textbf{a}}$.
\begin{thm}
\label{thm:suff}
$S_{\textbf{a}}$ is $C^0$-convergent if $S_{\textbf{b}}$ is contractive, i.e. $S_{\textbf{b}}$ maps any initial sequence of control points to $0$ upon iteration. 
\end{thm}
Note that contractivity of $S_{\textbf{b}}$ can be checked by computing if $\|S_{\textbf{b}}\|<1$ where 
\benn
\|S_{\textbf{b}}\|=\max\left\{\sum_j|b_{2j}|,\sum_j|b_{2j+1}|\right\}
\eenn
This reduces the question of convergence to a direct computation. Theorem \ref{thm:suff} can be improved to an ``if and only if'' statement in the case of asymptotic equivalence of the operators $S_{\textbf{a}}$ and $S_{\textbf{b}}$ (see \cite{2} for details). The easiest way to improve the level of differentiability is given in the next theorem. 
\begin{thm}
\label{thm:smooth}
Let $s_{\textbf{a}}(z)=\frac{(1+z)}{2}s_{\textbf{q}}(z)$. If $S_{\textbf{q}}$ is $C^k$-convergent then $S_{\textbf{a}}$ is $C^{k+1}$-convergent.
\end{thm}
Hence we can try to divide out $\frac{(1+z)}{2}$ factors from the z-transform and check the resulting scheme for convergence. Conversely we can improve the level of smoothness of a convergent scheme by multiplying the z-transform by $\frac{(1+z)}{2}$ factors. Note that the correspondence (\ref{eq:ztrans}) provides the corresponding mask $\textbf{a}$. Expositions and proofs of the previous results are found in \cite{1}, \cite{2} and \cite{3}. 

\section{Complex Eigenvalues: An Example and Minimum Width}

An alternative way of analyzing convergence is provided by setting up a matrix $A$ associated to $S_{\textbf{a}}$. Note that the refinement rules (\ref{eq:rules}) require an infinite matrix but to investigate smoothness near a given point it suffices to perform eigenanalysis on a finite matrix $A$ called the local subdivision matrix. Eigenanalysis provides an upper bound on the level of derivative continuity. If the width of $S_{\textbf{a}}$ is $n$ then $A\in\R^{n\times n}$. For a more detailed dsicussion of eigenanalysis consider \cite{3} or \cite{5}. \\

We ask the question if the matrix $A$ can have complex eigenvalues for a convergent scheme $S_{\textbf{a}}$. Reviewing many well-known subdivision schemes for curves the surprising result is that all classical schemes such as B-spline schemes or the four-point scheme (see \cite{3} or \cite{5}) do not have complex eigenvalues. A priori there is no reason for a matrix with real entries not to have complex eigenvalues. We remark that complex eigenvalues have been observed in the analysis of some subdivision schemes for surfaces such as the $\sqrt{3}$-scheme (see \cite{4}). For curves no complex eigenvalue scheme seems to be known. Therefore we ask three questions:

\begin{enumerate}
 \item Does there exist a convergent binary subdivision scheme with palindromic mask having complex eigenvalues? (Answer: Yes!)
 \item What is the minimum width of such a scheme?
 \item Why did complex eigenvalue schemes not appear ``naturally'' in computer-aided geometric design for curves?  
\end{enumerate}

\begin{prop}
There does not exist a binary subdivision scheme $S_{\textbf{a}}$ with width less than or equal to $5$ and complex eigenvalues in the local subdivision matrix $A$. 
\end{prop}
\begin{proof}
We begin with the case $n=5$. First note that since $s_{\textbf{a}}(z)=a_2z^2+a_1z+a_0+a_1z^{-1}+a_2z^{-2}$ we get using Theorem \ref{thm:nec} that $2=s_{\textbf{a}}(1)=2a_2+2a_1+a_0$ and $0=s_{\textbf{a}}(-1)=2a_2-2a_1+a_0$ for a convergent scheme. This implies the conditions:
\benn
a_0=1-2a_2\quad \text{ and } \quad a_1=\frac12
\eenn
Therefore the local subdivision matrix $A$ is given by 
\benn
A=\left( \begin{array}{cccccc}
 a & 1-2a & a & 0 & 0 \\
 0 & 1/2 & 1/2 & 0 & 0 \\
 0 & a & 1-2a & a & 0 \\
 0 & 0 & 1/2 & 1/2 & 0 \\
 0 & 0 & a & 1-2a & a \\
\end{array} \right)
\eenn
which has eigenvalues $\{1,\frac12,\frac12-2a,a,a\}$ and since $a\in\R$ we conclude that no convergent scheme with a mask of width $5$ can have complex eigenvalues. The cases $n\leq4$ are dealt with in the same way and all matrices have real eigenvalues in these cases.
\end{proof}

Note that the previous analysis fails for the case $n=6$. We show now that it is possible to find a convergent scheme of width $6$ with ``simple'' coefficients having a small common denominator.

\begin{prop}
\label{prop:ex}
The mask $\textbf{a}=\{a_{-2}=-\frac{1}{10},a_{-1}=\frac{3}{10},a_0=\frac45,a_1=\frac45,a_2=\frac{3}{10},a_2=-\frac{1}{10}\}$ defines a $C^0$-convergent binary scheme $S_{\textbf{a}}$ with complex eigenvalues in the local subdivision matrix $A$.
\end{prop}  
\begin{proof}
Since $s_{\textbf{a}}(z)=(1+z)\left(-\frac{z^2}{10}+\frac{2z}{5}+\frac{2}{5}+\frac{2}{5z}-\frac{1}{10z^2}\right):=(1+z)s_{\textbf{b}}(z)$ we find $\|S_{\textbf{b}}(z)\|=\frac45<1$ so that $S_{\textbf{a}}$ is $C^0$-convergent by Theorem \ref{thm:suff}. The local subdivision matrix is given by
\benn
A=\left( \begin{array}{cccccc}
 -1/10 & 4/5 & 3/10 & 0 & 0 & 0\\
 0 & 3/10 & 4/5 & -1/10 & 0 & 0\\
 0 & -1/10 & 4/5 & 3/10 & 0 & 0\\
 0 & 0 & 3/10 & 4/5 & -1/10 & 0\\
 0 & 0 & -1/10 & 4/5 & 3/10 & 0\\
 0 & 0 & 0 & 3/10 & 4/5 & -1/10\\
\end{array} \right)
\eenn
which has eigenvalues $\{-\frac{1}{10},-\frac{1}{10},\frac25,1,\frac15\left(2-i\sqrt2\right),\frac15\left(2+i\sqrt2\right)\}$ so that the result is proved.
\end{proof}
This means that complex eigenvalues impose conditions on the width of a mask. It is in general desirable to have masks with relatively small size to guarantee locality, i.e. moving a control point does not affect parts of the curve far away from this point. Many classical schemes have width between $6$ and $10$. Therefore we can conclude that there is no algebraic reason why we have not yet seen a complex eigenvalue scheme in practice.\\

We remark that we cannot use Theorem \ref{thm:smooth} to find a $C^1$-convergent scheme with width $6$. Since $s_{\textbf{a}}(z)=\frac{(1+z)}{2}\left(2az^2+2(b-a)z+2-4b+\frac{2(b-a)}{z}+\frac{2a}{z^2}\right)= \frac{(1+z)}{2}s_{\textbf{q}}(z)$ using Theorem \ref{thm:nec} we get the difference scheme $S_{\textbf{q}}$. Using Theorem \ref{thm:nec} on $S_{\textbf{q}}$ gives
\benn
0=s_{\textbf{q}}(-1)=2(1+4a-4b)\quad \Rightarrow \quad b=\frac14+a
\eenn  
This condition implies that the eigenvalues of the local subdivision matrix $A$ are real. But obviously we can use Theorem \ref{thm:nec} to construct a scheme with higher smoothness out of the complex eigenvalue scheme we presented by multiplying the z-transform with $\frac{(1+z)}{2}$ factors. For example we get from the scheme given in Proposition \ref{prop:ex} the $C^1$ scheme
\benn
\left\{a_{-3}=-\frac{1}{20},a_{-2}=\frac{1}{10},a_{-1}=\frac{11}{20},a_0=\frac45,a_1=\frac{11}{20},a_2=\frac{1}{10},a_3=-\frac{1}{20}\right\}
\eenn
It again has a pair of complex conjugate eigenvalues.

\section{Numerical Comparison: Complex vs. Real Eigenvalues}

Our goal is to answer the last of our three questions and we start by comparing the results of several schemes:

\begin{table}[htbp]
	\centering
	\label{tab:para}
	\begin{tabular}{|c|l|c|}
	\hline
	 & Mask & Smoothness\\
	\hline\hline
  (a) & $\left\{a_{-2}=-1/10,a_{-1}=3/10,a_0=4/5,a_1=4/5,a_2=3/10,a_2=-1/10\right\}$ & $C^0$ \\
  (b) & $\left\{a_{-3}=-1/20,a_{-2}=1/10,a_{-1}=11/20,a_0=4/5,a_1=11/20,a_2=1/10,a_3=-1/20\right\}$ & $C^1$ \\
  (c) & $\left\{a_{-1}=1/2,a_0=1,a_1=1/2\right\}$ & $C^0$ \\
  (d) & $\left\{a_{-2}=1/8,a_{-1}=4/8,a_0=6/8,a_1=4/8,a_2=1/8\right\}$ & $C^2$ \\
	\hline
	\end{tabular}
\end{table}

\begin{figure}[htbp]
	\centering
		\includegraphics[width=0.80\textwidth]{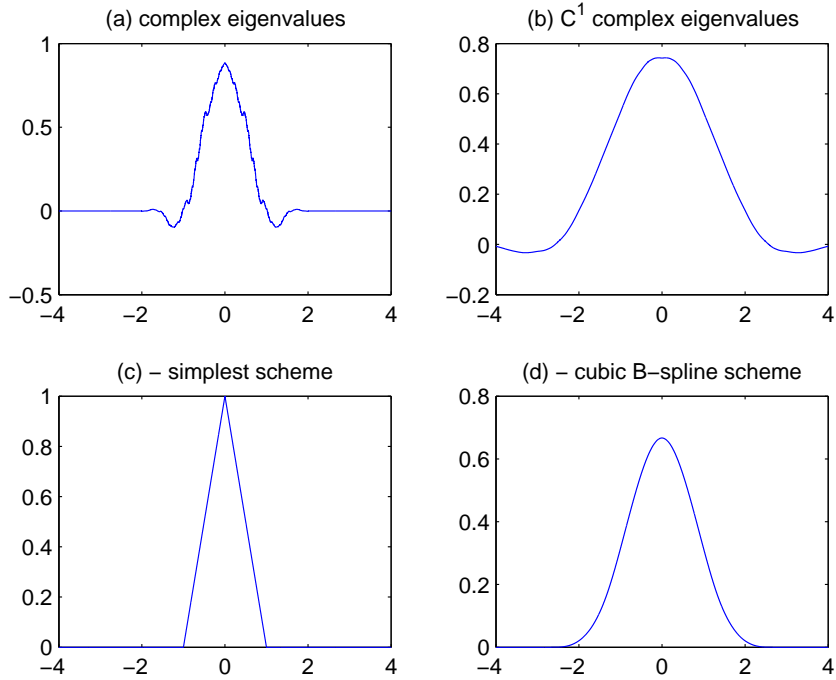}
	\caption{(a)-(b) complex eigenvalue schemes; (c)-(d) two ``standard'' schemes.} 
	\label{fig:fig1}
\end{figure}

Scheme (d) is derived from cubic B-splines and scheme (c) is known as ``simplest'' or ``two-point'' scheme; it is easy to check that (c)-(d) have only real eigenvalues. We compare the schemes with the most important test sequence of control points given by
\benn
\textbf{P}^0=\{0,0,0,0,1,0,0,0,0\}
\eenn
on the integer mesh $N^0$ on $[-4,4]$. The reason for its importance is that all convergence properties can be derived by investigating this sequence and using linearity of the scheme (see \cite{3}). The schemes were iterated 10 steps and linear interpolation was used to display the function constructed from $\textbf{P}^{10}$. The results are shown in Figure \ref{fig:fig1}.\\

The cubic B-spline scheme is $C^2$ and produces the expected smooth curve. The simplest scheme produces the classical $C^0$ tent function. For the complex eigenvalue scheme (a) we find a $C^0$ curve, which is neither smooth nor contains clear connecting points between several parts of the curve like displayed for the tent function. Instead we see small oscillations or ``ripples''. The complex eigenvalue scheme (b) gives a $C^1$ curve with two small negative bumps outside $[-2,2]$. In computer-aided geometric design undesired ripples or bumps of the limiting curve are usually called artefacts. In the last two decades the focus has been to find schemes which produce smooth curves or curves with well-defined sharp transitions without artefacts. This is one possible explanation why practioners might have overlooked or dismissed schemes with complex eigenvalues. In the next section we show why artefacts are expected to appear.

\section{A Geometric Explanation using Dynamical Systems} 

We aim to explain why we expect artefacts in every subdivision scheme for curves with complex eigenvalues. Fix a convergent scheme $S_{\textbf{a}}$ with one pair of complex conjugate eigenvalues. Fix a mesh point $x_{k_0}\in N^{k_{0}}$. Consider the $n$ control points in $\textbf{P}^{k_0}$ attached to mesh points closest to $x_{k_0}$. By shifting the indices we can assume without loss of generality that these points are given by $P^{k_0}_i$ for $i\in\{1,2,\ldots,n\}$. Define a vector 
\be
\label{eq:ic}
v_{k_0}=(P^{k_0}_1,P^{k_0}_2,\ldots,P^{k_0}_n)
\ee 
This process can be continued for the n-points closest to $x_{k_0}$ in $\textbf{P}^{k_0+1}$, $\textbf{P}^{k_0+2}$, etc. Note that the local subdivision matrix $A$ associated to $S_{\textbf{a}}$ produces this sequence of vectors under iteration:
\benn
v_{k+1}=Av_k \qquad \text{for $k\geq k_0$ and $v_{k_0}$ given by (\ref{eq:ic})}
\eenn
This means we are dealing with a particularly simple discrete dynamical system defined by the map $A:\R^n\ra \R^n$. It is known from eigenanalysis that the eigenvalues $\mu_i$ of $A$ must obey the restriction $\mu_1=1$ and $|\mu_i|<1$ for $i\in\{2,3,\ldots,n\}$ (see \cite{5}). Furthermore it is immediate by the convergence of $S_{\textbf{a}}$ that
\benn
 A v_k=v_{k+1}\ra \bar{v} \qquad \text{ as $k\ra \I$ for some vector vector $\bar{v}$ with identical components.}
\eenn
In fact we find that $A\bar{v}=\bar{v}$ so that $\bar{v}$ is a fixed point. Since we have at least one pair of complex conjugate eigenvalues there exists a change of basis so that $A$ has a two-dimensional eigenspace $R$ in which $A$ is a rotation-scaling with scaling factor $\lambda<1$. The scaling factor is given by the real part of the pair of complex conjugate eigenvalues. Due to the rotational component the convergence to $\bar{v}$ is not monotone. This lack of monotonicity is reflected in the different behaviour of the complex eigenvalue  schemes in comparison to some of the classical schemes.\\

In our examples of complex eigenvalues there is another reason for non-monotone convergence caused by the two real negative eigenvalues (see Proposition \ref{prop:ex}). This gives alternating convergence of two components of $v_k$ to $\bar{v}$. This alternating convergence is unavoidable for complex eigenvalue schemes with width $6$ as the next proposition shows.

\begin{prop}
Every binary complex eigenvalue subvision scheme with palindromic mask of width 6 has at least 2 eigenvalues with negative real part. 
\end{prop} 

\begin{proof}
The local subdivision matrix $A$ is given by
\benn
A=\left( \begin{array}{cccccc}
 a & c & b & 0 & 0 & 0\\
 0 & b & c & a & 0 & 0\\
 0 & a & c & b & 0 & 0\\
 0 & 0 & b & c & a & 0\\
 0 & 0 & a & c & b & 0\\
 0 & 0 & 0 & b & c & a\\
\end{array} \right)
\eenn
Using the condition $a+b+c=1$ we first eliminate $c$ and then compute the eigenvalues explicitly. They are given by:
\benn
\mu_1=1, \quad \mu_{2,3}=a, \quad \mu_4=b-a, \quad \mu_{5,6}=1-a-b\pm\sqrt{1+2a-7a^2-6b+2ab+9b^2} 
\eenn
For the eigenvalues to be complex we get that the terms inside the square-root for $\mu_{5,6}$ must be negative. The inequality to be satisfied is:
\benn
a<\frac{1+b-2\sqrt{2(1-5b+8b^2)}}{7}
\eenn
But we can calculate directly that
\benn
1+b-2\sqrt{2(1-5b+8b^2)}\leq 0\qquad  \text{for $b\in \R$}
\eenn
Therefore we can conclude that $\mu_{2,3}=a<0$.
\end{proof}

We leave it as an open question whether any binary complex eigenvalue scheme with palindromic mask must have eigenvalues with negative real part. Hence it might be possible to avoid alternating convergence for a complex eigenvalue scheme but non-monotonicity can obviously not be avoided. Hence we expect artefacts to appear.

\section{Conclusion}

We have demonstrated that subdivision schemes for curves can have complex eigenvalues. We have shown that the minimum width of a scheme with complex eigenvalues is $6$. Furthermore we defined a scheme achieving this minimum with simple coefficients. Numerical comparison of complex eigenvalue schemes with two well-known schemes showed that the main reason why complex eigenvalue schemes have been overlooked or disregarded is the appearance of ripples or artefacts. They appear due to the lack of monotonicity of the dynamical system defined by iteration of the local subdivision matrix $A$.\\

We recommend to consider the use of complex eigenvalue schemes in computer-aided geometric design if there is a need for curves or objects with rough surfaces. If the scaling factors of complex conjugate eigenvalues and the magnitude of real negative eigenvalues is small we expect a limiting curve with small artefacts. This can be used to scale how pronounced the artefacts should appear. Note that we can always construct a surface subdivision scheme from a complex eigenvalue curve scheme by using the tensor product of the mask. Therefore we can hope to use these schemes to define geometric objects which do not appear completely smooth or with sharp transitions and close a gap in the current practical applications of subdivision schemes.

\end{document}